\documentstyle[12pt,epsf]{article}
\newlength{\pubnumber} 

\catcode`\@=11
\@addtoreset{equation}{section}

\def\section{\@startsection{section}{1}{\z@}{3.5ex plus 1ex minus .2ex}
 {2.3ex plus .2ex}{\large\bf}}
\def\subsection{\@startsection{subsection}{2}{\z@}{2.3ex plus .2ex}
 {2.3ex plus .2ex}{\bf}}

\begin{document}

\begin{titlepage}
\samepage{
\setcounter{page}{0}
\rightline{TPI--MINN--00/42}
\rightline{UMN--TH--1918--00}
\rightline{OUTP--00--37P}
\rightline{\tt hep-ph/0008???}
\rightline{August 2000}
\vfill
\begin{center}
 {\Large \bf Self--Interacting Dark Matter from the \\
Hidden Heterotic--String Sector}
\vfill
\vspace{.25in}
 {\large Alon E. Faraggi$^{1,2}$ and Maxim Pospelov$^2$\\}
\vspace{.25in}
 {\it  $^1$Theoretical Physics Department, 
		University of Oxford, Oxford OX1 3NP, UK\\}
\vspace{.2in}
 {\it  $^2$Department of Physics,
              University of Minnesota, Minneapolis, MN  55455, USA\\}
\end{center}
\vfill
\begin{abstract}
  {\rm
It has been suggested recently that self--interacting dark matter
fits better the observational characteristics of galaxy dynamics. 
We propose that the self--interacting dark matter is composed from the 
glueballs of the hidden sector non--Abelian gauge group,
while the hidden matter states exist in vector--like 
representation and decouple from the light spectrum. It is shown that 
these glueballs are semi--stable with the life--time  
larger than the present age of the Universe, if their mass is 1 GeV or less. 
The constraint on their abundance today suggests that the energy 
was stored in the hidden sector soon after inflation. This imposes an 
upper limit on the reheating temperature. We further study the naturalness 
of this scenario in the context of the free--fermionic string models and
point out a class of such models where the self--interacting dark matter from
the hidden sector is indeed plausible. 
}
\end{abstract}
\vfill
\smallskip}
\end{titlepage}

\setcounter{footnote}{0}

\def\beq{\begin{equation}}
\def\eeq{\end{equation}}
\def\beqn{\begin{eqnarray}}
\def\la{\label}
\def\eeqn{\end{eqnarray}}
\def\Tr{{\rm Tr}\,}
\def\KM{{Ka\v{c}-Moody}}

\def\ie{{\it i.e.}}
\def\etc{{\it etc}}
\def\eg{{\it e.g.}}
\def\half{{\textstyle{1\over 2}}}
\def\third{{\textstyle {1\over3}}}
\def\quarter{{\textstyle {1\over4}}}
\def\m{{\tt -}}
\def\p{{\tt +}}

\def\rep#1{{\bf {#1}}}
\def\slash#1{#1\hskip-6pt/\hskip6pt}
\def\slk{\slash{k}}
\def\GeV{\,{\rm GeV}}
\def\TeV{\,{\rm TeV}}
\def\y{\,{\rm y}}
\def\SM{Standard-Model }
\def\SUSY{supersymmetry }
\def\SSM{supersymmetric standard model}
\def\vev#1{\left\langle #1\right\rangle}
\def\l{\langle}
\def\r{\rangle}

\def\Htw{{\tilde H}}
\def\chibar{{\overline{\chi}}}
\def\qbar{{\overline{q}}}
\def\ibar{{\overline{\imath}}}
\def\jbar{{\overline{\jmath}}}
\def\Hbar{{\overline{H}}}
\def\Qbar{{\overline{Q}}}
\def\abar{{\overline{a}}}
\def\alphabar{{\overline{\alpha}}}
\def\betabar{{\overline{\beta}}}
\def\tautwo{{ \tau_2 }}
\def\calF{{\cal F}}
\def\calP{{\cal P}}
\def\calN{{\cal N}}
\def\smallmatrix#1#2#3#4{{ {{#1}~{#2}\choose{#3}~{#4}} }}
\def\bone{{\bf 1}}
\def\V{{\bf V}}
\def\N{{\bf N}}
\def\bQ{{\bf Q}}
\def\t#1#2{{ \Theta\left\lbrack \matrix{ {#1}\cr {#2}\cr }\right\rbrack }}
\def\C#1#2{{ C\left\lbrack \matrix{ {#1}\cr {#2}\cr }\right\rbrack }}
\def\tp#1#2{{ \Theta'\left\lbrack \matrix{ {#1}\cr {#2}\cr }\right\rbrack }}
\def\tpp#1#2{{ \Theta''\left\lbrack \matrix{ {#1}\cr {#2}\cr }\right\rbrack }}
\def\ul#1{$\underline{#1}$}
\def\bE#1{{E^{(#1)}}}
\def\IZ{\relax{\bf Z}}\def\IC{\relax{\bf C}}
\def\IR{\relax{\rm I\kern-.18em R}}
\def\lamb{\lambda}
\def\fc#1#2{{#1\over#2}}
\def\hx#1{{\hat{#1}}}
\def\Gh{\hat{\Gamma}}
\def\subsubsec#1{\noindent {\it #1} \br}
\def\WP{{\bf WP}}
\def\gn{\Gamma_0}
\def\bgn{{\bar \Gamma}_0}
\def\Ds{\Delta^\star}
\def\abstract#1{
\vskip .5in\vfil\centerline
{\bf Abstract}\penalty1000
{{\smallskip\ifx\answ\bigans\leftskip 2pc \rightskip 2pc
\else\leftskip 5pc \rightskip 5pc\fi
\noindent\abstractfont \baselineskip=12pt
{#1} \smallskip}}
\penalty-1000}
\def\us#1{\underline{#1}}
\def\hth/#1#2#3#4#5#6#7{{\tt hep-th/#1#2#3#4#5#6#7}}
\def\nup#1({Nucl.\ Phys.\ $\us {B#1}$\ (}
\def\plt#1({Phys.\ Lett.\ $\us  {B#1}$\ (}
\def\cmp#1({Commun.\ Math.\ Phys.\ $\us  {#1}$\ (}
\def\prp#1({Phys.\ Rep.\ $\us  {#1}$\ (}
\def\prl#1({Phys.\ Rev.\ Lett.\ $\us  {#1}$\ (}
\def\prv#1({Phys.\ Rev.\ $\us  {#1}$\ (}
\def\mpl#1({Mod.\ Phys.\ Let.\ $\us  {A#1}$\ (}
\def\ijmp#1({Int.\ J.\ Mod.\ Phys.\ $\us{A#1}$\ (}
\def\br{\hfill\break}\def\ni{\noindent}
\def\mbr{\hfill\break\vskip 0.2cm}
\def\cx#1{{\cal #1}}\def\al{\alpha}\def\IP{{\bf P}}
\def\ov#1#2{{#1 \over #2}}
\def\b{{\bf b}}
\def\S{{\bf S}}
\def\X{{\bf X}}
\def\I{{\bf I}}
\def\mb{{\mathbf b}}
\def\mS{{\mathbf S}}
\def\mX{{\mathbf X}}
\def\mI{{\mathbf I}}
\def\balpha{{\mathbf \alpha}}
\def\bbeta{{\mathbf \beta}}
\def\bgamma{{\mathbf \gamma}}
\def\bxi{{\mathbf \xi}}
 
\def\ul#1{$\underline{#1}$}
\def\bE#1{{E^{(#1)}}}
\def\IZ{\relax{\bf Z}}\def\IC{\relax{\bf C}}
\def\IR{\relax{\rm I\kern-.18em R}}
\def\lam{\lambda}
\def\fc#1#2{{#1\over#2}}
\def\hx#1{{\hat{#1}}}
\def\Gh{\hat{\Gamma}}
\def\subsubsec#1{\noindent {\it #1} \br}
\def\WP{{\bf WP}}
\def\gn{\Gamma_0}
\def\bgn{{\bar \Gamma}_0}
\def\Ds{\Delta^\star}
\def\abstract#1{
\vskip .5in\vfil\centerline
{\bf Abstract}\penalty1000
{{\smallskip\ifx\answ\bigans\leftskip 2pc \rightskip 2pc
\else\leftskip 5pc \rightskip 5pc\fi
\noindent\abstractfont \baselineskip=12pt
{#1} \smallskip}}
\penalty-1000}
\def\us#1{\underline{#1}}
\def\hth/#1#2#3#4#5#6#7{{\tt hep-th/#1#2#3#4#5#6#7}}
\def\nup#1({Nucl.\ Phys.\ $\us {B#1}$\ (}
\def\plt#1({Phys.\ Lett.\ $\us  {B#1}$\ (}
\def\cmp#1({Commun.\ Math.\ Phys.\ $\us  {#1}$\ (}
\def\prp#1({Phys.\ Rep.\ $\us  {#1}$\ (}
\def\prl#1({Phys.\ Rev.\ Lett.\ $\us  {#1}$\ (}
\def\prv#1({Phys.\ Rev.\ $\us  {#1}$\ (}
\def\mpl#1({Mod.\ Phys.\ Let.\ $\us  {A#1}$\ (}
\def\ijmp#1({Int.\ J.\ Mod.\ Phys.\ $\us{A#1}$\ (}
\def\br{\hfill\break}\def\ni{\noindent}
\def\mbr{\hfill\break\vskip 0.2cm}
\def\cx#1{{\cal #1}}\def\al{\alpha}\def\IP{{\bf P}}
\def\ov#1#2{{#1 \over #2}}
\def\ga{\mathrel{\raise.3ex\hbox{$>$\kern-.75em\lower1ex\hbox{$\sim$}}}}
\def\la{\mathrel{\raise.3ex\hbox{$<$\kern-.75em\lower1ex\hbox{$\sim$}}}}
\def\deg{\hbox{${}^\circ$}}     


\def\inbar{\,\vrule height1.5ex width.4pt depth0pt}

\def\IC{\relax\hbox{$\inbar\kern-.3em{\rm C}$}}
\def\IQ{\relax\hbox{$\inbar\kern-.3em{\rm Q}$}}
\def\IR{\relax{\rm I\kern-.18em R}}
 \font\cmss=cmss10 \font\cmsss=cmss10 at 7pt
\def\IZ{\relax\ifmmode\mathchoice
 {\hbox{\cmss Z\kern-.4em Z}}{\hbox{\cmss Z\kern-.4em Z}}
 {\lower.9pt\hbox{\cmsss Z\kern-.4em Z}}
 {\lower1.2pt\hbox{\cmsss Z\kern-.4em Z}}\else{\cmss Z\kern-.4em Z}\fi}

\def\AEF{A.E. Faraggi}
\def\KRD{K.R. Dienes}
\def\JMR{J. March-Russell}
\def\MEP{M.E. Pospelov}
\def\NPB#1#2#3{{\it Nucl.\ Phys.}\/ {\bf B#1} (#2) #3}
\def\PLB#1#2#3{{\it Phys.\ Lett.}\/ {\bf B#1} (#2) #3}
\def\PRD#1#2#3{{\it Phys.\ Rev.}\/ {\bf D#1} (#2) #3}
\def\PRL#1#2#3{{\it Phys.\ Rev.\ Lett.}\/ {\bf #1} (#2) #3}
\def\PRT#1#2#3{{\it Phys.\ Rep.}\/ {\bf#1} (#2) #3}
\def\MODA#1#2#3{{\it Mod.\ Phys.\ Lett.}\/ {\bf A#1} (#2) #3}
\def\IJMP#1#2#3{{\it Int.\ J.\ Mod.\ Phys.}\/ {\bf A#1} (#2) #3}
\def\nuvc#1#2#3{{\it Nuovo Cimento}\/ {\bf #1A} (#2) #3}
\def\etal{{\it et al,\/}\ }

\hyphenation{su-per-sym-met-ric non-su-per-sym-met-ric}
\hyphenation{space-time-super-sym-met-ric}
\hyphenation{mod-u-lar mod-u-lar--in-var-i-ant}
\newcommand{\be}{\begin{equation}}
\newcommand{\ee}{\end{equation}}
\newcommand{\als}{\mbox{$\alpha_{s}$}}
\newcommand{\s}{\mbox{$\sigma$}}
\newcommand{\lm}{\mbox{$\mbox{ln}(1/\alpha)$}}
\newcommand{\bi}[1]{\bibitem{#1}}
\newcommand{\fr}[2]{\frac{#1}{#2}}
\newcommand{\sv}{\mbox{$\vec{\sigma}$}}
\newcommand{\gm}{\mbox{$\gamma_{\mu}$}}
\newcommand{\Pm}{\mbox{$P_{\mu}$}}
\newcommand{\Pn}{\mbox{$P_{\nu}$}}
\newcommand{\Pa}{\mbox{$P_{\alpha}$}}
\newcommand{\ph}{\mbox{$\hat{p}$}}
\newcommand{\Ph}{\mbox{$\hat{P}$}}
\newcommand{\qh}{\mbox{$\hat{q}$}}
\newcommand{\kh}{\mbox{$\hat{k}$}}
\newcommand{\Le}{\mbox{$\fr{1+\gamma_5}{2}$}}
\newcommand{\R}{\mbox{$\fr{1-\gamma_5}{2}$}}
\newcommand{\GD}{\mbox{$\tilde{G}$}}
\newcommand{\gf}{\mbox{$\gamma_{5}$}}
\newcommand{\om}{\mbox{$\omega$}}
\newcommand{\Ima}{\mbox{Im}}
\newcommand{\Rea}{\mbox{Re}}


\setcounter{footnote}{0}
\section{Introduction}
Substantial experimental evidence indicates that most of the mass
in the universe is invisible. The determination of the nature of
this dark matter is one of the important challenges confronting
modern physics. The Cold Dark Matter (CDM) class of cosmological
models provides good description for a wide variety of observational
results, ranging from the early universe, probed via the microwave
background fluctuations to present day observations of galaxies and large
scales structure. Flat cosmological models with a mixture of
baryonic matter, cold matter and vacuum energy, can account for
almost all observations on scales $\ge$ 1 Mpc. More recently, 
improved observations and numerical simulations have enabled
comparison of CDM models to observations on galactic scales
of $\sim$ few kpc \cite{moore}. These studies reveal that the collisionless
CDM scenarios, which predict halo density profiles that are singular at the
center, are in apparent contradiction with observations, which
indicate uniform density cores. This conflict prompted 
Spergel and Steinhardt \cite{sps} to propose that the cold dark matter
is self--interacting with a large scattering cross section
but negligible annihilation or dissipation. 
The key feature of this proposal is that the 
mean free path of the self--interacting dark matter candidate
should be $1~{\rm kpc}<\lambda_{\rm free~path}<1~{\rm Mpc}$. 
The effect of the self--interaction
would smooth out the central profiles and suppress the number of
satellite galaxies, hence improving the agreement with
observations \cite{sps}. 

In view of these developments, it is prudent to explore
possible particle dark matter candidates that possess the
required properties. One can imagine that to devise a 
particle with these desired characteristics is by no
means too difficult. It is therefore essential to examine whether
a particle with the coveted virtues, can be motivated 
in a larger context. This is for example the situation
in the case of the very well motivated Cold Dark Matter candidates,
like the neutralino and the axion, which are motivated,
respectively, by supersymmetry and the strong CP problem. 

In this paper we therefore study self--interacting dark matter which is
motivated from string theory. One particular class of string
motivated dark matter candidates are the strongly interacting 
states, the uniton and the sexton, which were proposed
in ref. \cite{ccf}. However, in ref. \cite{fop}
it was shown that such states would accumulate in the
center of the sun and the earth and would subsequently
annihilate into energetic neutrinos at an unacceptable rate. 
Moreover,  even though their self-interaction is sufficiently strong and the 
scattering cross-section is of the order of hadron-hadron scattering 
cross-section, their masses are expected to be quite large, which translates 
into a low number density and very large $\lambda_{\rm free~path}$. 
What about other strongly interacting candidates, such as gluino LSP
scenario, advocated in Refs. \cite{F,cfk,FSS}? If the gluino is heavier than 
1 GeV, this scenario again could be excluded from the indirect searches of 
the dark matter via the flux of energetic neutrinos. It seems, 
however, that if the mass of gluino is really low, {\it i.e.} $\leq 1$ GeV,
it will
not produce sufficiently energetic neutrinos, the indirect constraints do not 
apply, and, as was argued in Ref. \cite{FSS}, direct searches are also not 
sensitive to gluino--containing hadrons, as they will be considerably slowed 
down before reaching an underground detector. 

Here we study a different type of self--interacting
dark matter candidate which comes from the hidden strongly-interacting
sector of the theory, reminiscent to what was considered in Ref. \cite{Okun}
twenty years ago. The existence of such a  hidden
sector is a generic consequence of the string theory. 
In particular we examine 
the case that the lightest hidden sector state
is a stable glueball of a non--Abelian hidden 
gauge group, which arise from the hidden string sector, 
and interacts with the Standard Model states only via
hidden sector heavy matter states. As is frequently the case in
semi--realistic heterotic--string derived models, 
the hidden sector matter states are charged with
respect to horizontal $U(1)$ symmetries which are
broken near the string scale, and under which also the
Standard Model states are charged. Therefore, the 
lightest hidden sector glueball states are strongly 
interacting among themselves and are very weakly interacting
with the Standard Model states, where interactions are
mediated by higher dimensional operators, which are suppressed
by inverse powers of the string scale. Generically, these hidden sector 
glueballs are metastable, with strong dependence of the lifetime on
the condensation scale.  

In view of recent years progress in string theory, 
one must address the issue of what is the appropriate
string scale to use. We pursue the minimalist approach
which assumes the big desert scenario as suggested
by the Standard Model multiplet structure and 
grand unification. The relevant framework is
that of the heterotic--string and the string scale
is of the order $10^{16-17}~{\rm GeV}$. We further 
assume that the Standard Model gauge couplings as
well as those of the hidden sector unify near the
string scale, and are of the order extracted by
the standard extrapolation of the MSSM gauge couplings. 
We then examine the constraints that are imposed on the
possible non--Abelian hidden gauge groups by the results
indicated by Spergel and Steinhardt and by the 
requirement that the lightest stable hidden state
constitute the dark matter. We find that this
set of assumptions rather tightly constrain the possible hidden sectors
that can produce the desired characteristics.
We examine the possible emergence of stable hidden states
with these virtues from heterotic--theory. 
We show that a class of three generation free fermionic 
heterotic string models can in fact produce a hidden
sector with the desired properties. The basic features
that are needed, and which are reproduced in this class
of string models, is a non--Abelian hidden gauge group 
with a small gauge content and that the hidden
matter appears in vector--like representations and
can decouple at a sufficiently high scale. 

\section{Hidden glueballs dark matter}
Spergel and Steinhardt propose that the dark matter particles should have
a mean free path, $1{\rm kpc}\le\lambda_{\rm free~path}\le1{\rm Mpc}$.
For a particle with mass $m_x$ this implies an elastic scattering cross
section of 
\beq
\sigma_{XX}=8.1\cdot 10^{-25}{\rm cm}^2\left({m_x\over{\rm GeV}}\right)
\left({\lambda\over{1~{\rm Mpc}}}\right)^{-1}
\label{sigmaxx}
\eeq
Assuming that the dark matter particle scatter through strong
interactions similar to hadronic scattering, the cross section is
approximately equal to the geometric cross section,
$\sigma\sim4\pi a^2$, where $a$ is the scattering length. 
Assuming $a\approx100fm_x^{-1}$ Spergel and Steinhardt obtain the 
estimate
\beq
m_x=4\left({\lambda\over{1{\rm Mpc}}}\right)^{1/3}f^{2/3}{\rm GeV}
\label{mx}
\eeq

Here we study the possibility that the self--interacting dark matter
(SIDM) comes from the hidden non--Abelian sector of the theory. 
The effective low--energy gauge symmetry is therefore
that of the Standard Model plus an hidden gauge group, 
which can be $SU(2)$, $SU(3)$ or another. However, as we
elaborate in section \ref{string}, inspired from the realistic
heterotic--string models, we assume that all the hidden matter
fields appear in vector--like representations, and can therefore
decouple from the light spectrum at a higher scale.  
In section \ref{rge} we will show that, assuming unification of
the Standard Model as well as the hidden sector couplings, 
strongly constrains the allowed possibilities. If the additional
hidden groups are not Higgsed, then at some scale $\Lambda_h$, 
because of asymptotic freedom, the hidden sector gauge coupling,
$g_h$, becomes large and the hidden gauge group will be in the 
confining regime. Necessarily it will develop a mass gap in the 
spectrum $\sim O(\Lambda_h)$ and the lowest glueball--like
state will be stable. These particles will be strongly interacting 
among themselves and ``almost'' non--interacting with the Standard Model
particles. 

The lightest hidden sector state is non--interacting with the
Standard Model states up to higher dimensional operators which
are generated at the radiative level. These lowest order terms 
are of the form
\beq
{\cal L}_{\rm eff}=C_6{{(H^\dagger H)(G_{\mu\nu}^aG_{\mu\nu}^a)}\over{M_S^2}}
+\sum_{\rm SM}C^i_8{{(G_{\mu\nu}^aG_{\mu\nu}^a)
(F_{\mu\nu}F_{\mu\nu})}\over{M_S^4}}+
\sum_{\rm SM}\tilde C^i_8{{(G_{\mu\nu}^a\tilde G_{\mu\nu}^a)
(F_{\mu\nu}\tilde F_{\mu\nu})}\over{M_S^4}}+...
\label{hdoperators}
\eeq
Here $G_{\mu\nu}^a$ is the field strength of a non-Abelian gauge 
group from the hidden sector, $F_{\mu\nu}$ represents the field 
strength of the SM gauge groups ($U(1)$, $SU(2)$, or $SU(3)$). The 
summation runs over the SM gauge groups and ellipses stands for other possible 
operators. 
$C_6$, $C_8^i$, and $\tilde C_8^i$ are loop coefficients; $M_S$ is a heavy mass
scale, associated with the decoupling of the heavy
vector--like matter fields, which transform under the hidden
gauge group.  These fields are also charged under horizontal
$U(1)$ symmetries, under which also the Standard Model fields 
are charged. Note that the scale $M_S$ does not necessarily 
coincide with the string scale, although in the simplest 
scenarios that we consider we will assume that that is the case.
We remark that $C_6$ may have an additional suppression which
depends on the Yukawa couplings. Eq. (\ref{hdoperators}) is given in a 
non-supersymmetric form, and we note that the supersymmetric 
generalization is straightforward. 

It is now possible to make a rough 
estimate of the lifetime of the hidden glueball, which decay is induced by 
operators (\ref{hdoperators}).
Let us assume that the lightest glueball is a scalar. Then, the operator
$(G_{\mu\nu}^aG_{\mu\nu}^a)$ can annihilate an exotic
glueball (exoglueball) with the efficiency
\beq
<0|G_{\mu\nu}^aG_{\mu\nu}^a|{\rm exoglueball}>=f\Lambda_h^3\phi_h,
\label{efficiency}
\eeq
where $\phi$ is the wave--function of the glueball and $f$ is some
dimensionless coupling, presumably of the order one.
Thus, at $\Lambda_h$ and below the effective Lagrangian 
for the interaction of the exotic glueball with the 
Standard Model states becomes
\beq
{\cal L}=C_6f{{\Lambda_h^3}\over{M_S^2}}\phi(H^\dagger H)+
\sum C_8^if{{\Lambda_h^3}\over{M_S^4}}\phi({F}_{\mu\nu}F_{\mu\nu})
\label{ehdoperator}
\eeq
We first estimate the lifetime expected from the dimension
eight operator, involving SM SU(3) fields. This induces hadronic decays of 
exotic glueballs, with the probability per unit time given by 
\beq
\Gamma={2m_\phi^3\over\pi}\left(C_8f{\Lambda_h^3\over{M_S^4}}\right)^2
\sim\Lambda_h\left({\Lambda_h\over{M_S}}\right)^8,
\label{d8gamma}\
\eeq
where we omitted (likely small) numerical coefficients and used the fact that 
$m_\phi \sim \Lambda_h$.
Taking $M_S\approx 10^{16}{\rm GeV}$ and $\Gamma<1/\tau_{\rm universe}=
1/(10^{10}{\rm yr})$, we obtain the following condition
\beq
\Lambda_h< 3\cdot 10^9{\rm GeV}
\label{estimatelambdah8}
\eeq
Thus, for a reasonable $\Lambda_h$ the decay rate is smaller
than the inverse lifetime of the universe.

We next turn to estimate the lifetime expected from the dimension six
operator. The decay is mediated by the virtual Higgs particle, decaying 
into all possible channels. 
Motivated by Spergel and Steinhardt's proposal, we consider $m_\phi$ in 
the ballpark of 1 GeV, which means that the decay width is saturated by 
hadronic channels and the effective Lagrangian can be 
written in the following form: 
\beq
{\cal L}_{\rm eff}=C_6f{{\Lambda_h^3}\over{M_{Higgs}^2 M_S^2}}
{3\alpha_3\over 8\pi}(F^a_{\mu\nu}F^a_{\mu\nu}),
\label{d6estimate}
\eeq
where $F_{\mu\nu}^a$ now is the gluon $SU(3)$ field strength. 
Using same formulae as before, we get
\beq
\Gamma_\phi=(C_6f)^2{2m_\phi^3\over\pi}\left(
{3\alpha_3\over 8\pi }\right)^2
\left({{\Lambda_h^3}\over{M_S^2M_{Higgs}^2}}
\right)^2 \sim 10^{-2}
{{\Lambda_h^9}\over {M_S^4M_{Higgs}^4}}
\label{estimatelambdah6}
\eeq
The condition $\Gamma\cdot\tau_{\rm universe}<1$ yields, 
$\Lambda_h\le 10^5{\rm GeV}$, which is well above
the range for $m_x$, required by self--interacting dark matter scenario.

Thus we conclude, that {\em if} the requirement of 
strongly-interacting dark matter is satisfied, and the condensation scale 
$\Lambda_h$ is 
around 1 GeV or less, the lifetime of exotic glueballs is much larger than 
the present age of the Universe, and thus these glueballs might constitute
(or partially contribute to) the dark matter. 
Now we turn to analyzing possible scenarios which could give a required 
cosmological abundance of these glueballs close to $\Omega_\phi h^2 \sim 1$.

\section{Estimate of primordial abundance}

We start this section by noting that there are no ``natural'' reasons 
for the mass density of exotic glueballs to be of the order of the required 
dark matter density. In this respect the situation is quite different from
the neutralino LSP dark matter, where a correct abundance follows from the 
weak--scale annihilation cross section.  
In the case of exotic glueballs the low-energy annihilation
cross section into the SM particles is extremely low, exactly
for the same reasons that the decay width of an individual 
glueball turns out to be so small. When the annihilation cross section
is so small, there is always a ``danger'' of overproducing these particles,
so that they overclose the Universe. 

It is possible to show on general grounds that the hidden sector 
cannot be in thermal equilibrium with observable matter.  In a deconfining 
phase the energy density of the exotic gluons scales
as $1/{R^4}$ with the size of the Universe $R$, and after 
the phase transition it scales as $1/R^3$. Thus, 
if at some point in the history of the Universe, 
the exotic gluons were in thermal equilibrium with normal matter,
they carried comparable energy density and  the confining phase
transition should have occurred around the same $R_{\rm eq}$ when 
the normal radiation--domination $\rightarrow$ matter--domination phase 
transition occurs. 
That is, around $T_{\rm normal~matter}\sim5{\rm eV}$
and long after nucleosynthesis. It implies that during the 
nucleosynthesis there were $2(N_C^2-1)$ additional massless
degrees of freedom, associated with exotic gluons. 
For $SU(2)$ hidden gauge group we therefore get additional 6
degrees of freedom, seemingly in disagreement with the nucleosynthesis 
constraints. If the deconfining--confining phase transition 
occurred before nucleosynthesis, the energy density stored in exotic matter 
would have been much larger than observable matter density today. 
Therefore, the exotic gluons could not have been
in thermal equilibrium with ordinary matter.
Consequently, we have to come up with some mechanism responsible for 
their dilution and we assume that the exotic gluons
were produced after inflation with $T_{\rm Reating}<M_S$. 

In that case, several possibilities are open. Some portion of an 
inflaton could decay directly into the exotic gluons, a straightforward 
possibility, which requires a fine--tuning of the corresponding coupling. 
Another, more interesting, case is when the coupling of the inflaton to the 
exotic sector is nil, and the decay occurs predominantly into the observable 
sector. Due to the existence of small but non-vanishing  couplings between 
the two sectors, Eq. (\ref{hdoperators}), the high-energy scattering of 
visible sector particles would lead to the creation of hidden sector 
particles with the efficiency $C_n^2T_R^{2n+1}/M_S^{2n}$, where 
$n$ is 2 or 4, depending on 
the operator, and $T_R$ is 
the temperature of visible sector matter, which defines 
the energy density of the visible sector matter.
\beq
\rho_R={\pi^2\over{30}}g_*T_R^4
\label{normalro}
\eeq
Let us denote the exotic gluon radiation density by $\rho_g$ and write down a 
system of cosmological equations which would determine the evolution of the 
hot Universe: 
\beqn
H^2&=&{{8\pi}\over{3M_{\rm Pl}^2}}\left(\rho_R+\rho_g\right)\\
{{d\rho_R}\over{dt}}&=&-4H\rho_R-\alpha C_n^2{T_R^{2n+5}\over M_S^{2n}}\\
{{d\rho_g}\over{dt}}&=&-4H\rho_g+\alpha C_n^2{T_R^{2n+5}\over M_S^{2n}},
\label{ddt}
\eeqn
where $\alpha$ is some dimensionless constant, with no  parametric dependence 
on any of the relevant scales in the problem. 
These equation can be considerably simplified because $\rho_g$ gives a 
negligibly small contribution to the expansion rate of the hot universe, 
and away from dynamical equilibrium the dilution of $\rho_R$ due to the 
``leakage'' into the hidden sector can be safely neglected:
\beqn
H^2&=&{{8\pi}\over{3M_{\rm Pl}^2}}\rho_R,\\
{{d\rho_R}\over{dt}}&=&-4H\rho_R\\
{{d\rho_g}\over{dt}}&=&-4H\rho_g+\alpha C_n^2{T_R^{2n+5}\over M_S^2n}.
\label{aproddt}
\eeqn
These equations reduce the problem to finding the energy density 
$\rho_g$ as the function of ``external'' temperature $T_R$, 
$\rho_g(t(T_R))$.   
 Using the relation
\beq
t={1\over2}H^{-1}={1\over2}{1\over{\sqrt{{{\pi^2}\over{30}}g_{*{\rm SM}}}}}
{M_{\rm Pl}\over{T_R^2}}
\eeq
we obtain a simple differential equation for $\rho_g(T_R)$:
\beqn
{{d\rho_g}\over{dt}}=&-&T_R^3{{d\rho_g}\over{dT_R}}
{{\sqrt{{{\pi^2}\over{30}}g_{*{\rm SM}}}}\over M_{\rm Pl}}=\nonumber\\
&-&\sqrt{{{\pi^2}\over{30}}g_{*{\rm SM}}}{T_R^2\over M_{\rm Pl}}\rho_g+
\alpha C_n^2{T_R^{2n+5}\over{M_S^{2n}}}
\label{drhogdt}
\eeqn
Integrating this equation explicitly, and using 
the initial condition $\rho_g(T_{\rm RH})=0$, we arrive at
\beq
\rho_g(T_R)\sim{{\alpha C_n^2}\over
{\sqrt{{\pi^2\over{30}}g_{*{\rm SM}}}}}{M_{\rm Pl}\over
{M_S^{2n}}}{1\over{2n-1}}(T_{\rm RH}^{2n-1}-T_R^{2n-1})T_R^4
\label{rhogvstr}
\eeq
More accurate treatment would require some knowledge about reheating, as most 
of $\rho_g$ created at $T_R$ close to$ T_{RH}$.  
As one would naturally expect, soon after the reheating, the ratio of 
two energy densities becomes proportional to the dimensionless combination of 
the Plank scale, reheating temperature and $M_S$:
\beq
{\rho_g\over\rho_R} \sim {{M_{\rm Pl}T_{\rm RH}^{2n-1}}\over
{M_S^{2n}}}.
\eeq
After quick rescattering and thermalization, the exotic gluons will acquire 
their own temperature $T_g$, which 
will be lower than $T_R$, because we explicitly assume that 
${{M_{\rm Pl}T_{\rm RH}^{2n-1}}\over {M_S^{2n}}}<<1$. 
After that $T_g^4$ and $T_R^4$ both scale as
$1/R^4$, until $T_g$ cools down to $\Lambda_h$, where the confining phase 
transition occurs. From that moment the energy density, stored in 
exotic glueballs, scales as
$1/R^3$. After $T_R\approx5{\rm eV}\equiv T_{\rm EQ}$
both the normal matter and dark matter scale as $1/R^3$. 
This leads to the following final estimate of the abundance,
\beq
{{\Omega_{\rm dark}}\over{\Omega_{\rm baryon}}}\approx
{{T_R({\rm at}~T_g=\Lambda_h)}\over{T_{\rm EQ}}}
{{T_{\rm RH}^{2n-1}M_{\rm Pl}}\over{M_S^{2n}}}=
{\Lambda_h\over{T_{\rm EQ}}}\left(C_n^2
{{T_{\rm RH}^{2n-1}M_{\rm Pl}}\over{M_S^{2n}}}\right)^{3/4},
\label{omeoveome}
\eeq
which should be of order $O(10)$. This relation can viewed
as the condition on the reheating temperature, because
$\Lambda_h$ is more or less fixed by the requirement (2.2) 
and can be further related to $M_S$ via the renormalization 
group flow. Assuming $\Lambda_h\sim1{\rm GeV}$, $T_{\rm EQ}\sim
5{\rm eV}$, and $M_S\approx10^{16}{\rm GeV}$, we can get
$T_{\rm RH}\sim 10^{-4}M_S$ for the $n=2$ maximal strength operator,
and $T_{\rm RH}\sim10^{-2}M_S$ for $n=4$. This range for the reheating
temperature is quite reasonable. Scenarios with lower values of $T_{RH}$ 
would require a direct coupling between inflaton and the hidden sector 
to ensure sufficient abundance of exotic glueballs. 
On the other hand, scenarios with larger $T_{RH}$ in units of $M_S$ 
are excluded as in this case the energy density stored 
in exotic glueballs would overclose the universe. 

\section{Renormalization Group Analysis}\label{rge}

In the previous section we showed that, assuming $\Lambda_h\sim1{\rm GeV}$,
the exotic glueballs can account for the missing mass, and be in agreement
with nucleosynthesis constraints, provided that the reheating temperature
is of the order $O(10^{12-14}{\rm GeV})$. In this section we study
the evolution of the gauge couplings from the unification scale 
to the low scale. In this respect we examine which gauge groups,
and under what conditions, can produce the exotic glueballs
at a scale of the order $O(1){\rm GeV}$, as suggested
by the cosmological data.

As we discussed in the introduction, in order to make connection with 
realistic superstring models we assume the framework of the 
Supersymmetric Standard Model unification,
which indicates that the unification scale is of the order
$10^{16-17}{\rm GeV}$.  Consistently with this assumption, we
also assume the framework of the heterotic string, and consequently
that the observable and hidden gauge couplings unify at the unification scale.
With these assumptions, we extrapolate the hidden sector gauge
coupling from the high to low scales. We take the magnitude
of the hidden sector gauge coupling at the unification scale 
to be of the order that is expected by extrapolation of the 
Supersymmetric Standard Model couplings from the low scale 
to the unification scale. 
We divide our analysis into two
parts. We first assume that all the vector--like
matter states decouple at a scale which is identical with the $M_S$ scale.
Therefore below the $M_S$ scale only the gauge sector contributes
to the evolution of the gauge couplings. In the subsequent part
we assume that the vector--like matter states decouple 
at an intermediate scale $M_h$, which is below $M_S$. 
In the evolution of the couplings we assume a supersymmetric
spectrum.
In the next section we will examine 
the plausibility of obtaining the required particle 
content from realistic superstring models.

Assuming 
that there are no intermediate scales between $M_S$
and $\Lambda_h$, the scale at which the hidden $SU(n)$ gauge 
group becomes strongly interacting is given by, 
\beq
\Lambda_h=M_S{\rm Exp}{\left({{{2\pi}\over{({1\over2}N_f-3 N_C)}}
{{(1-\alpha_0)}\over\alpha_0}}\right)}, 
\label{lambdah}
\eeq
where $N_c=n$ is the number of colors,
$N_f$ is the number of spin 1/2 fundamental multiplets,
and $\alpha_0$ is the value of gauge coupling at the scale $M_S$.
Assuming that all matter states decouple at $M_S$ gives $N_f=0$. 
Then for the lowest possibility with $N_C=2$, and taking 
and $M_S\approx10^{16}{\rm GeV}$, $\Lambda_h\sim 1$ GeV, we obtain 
the required initial value of the gauge coupling, $\alpha_0\simeq 1/36$. 
This value of the coupling is 1.5 times smaller than conventional value
$\alpha_0\simeq 1/24$, suggested by the unification of coupling constants 
from the observable sector. If we take $1/24$ for the 
value of the coupling, the 
transition to the strong coupling regime for the hidden sector will occur at
$\Lambda_h\sim 3\cdot 10^{15}$ GeV, which is, clearly, too high a scale
to satisfy the strongly--interacting dark matter criterion.     

However, although this scenario is appealing in its
simplicity, a likely outcome suggested by
realistic string models is that the hidden matter 
states decouple at an intermediate energy scale which
is slightly or several orders of magnitude below
the string scale. We will discuss the relevant string 
framework in the subsequent section. Here we continue with
our qualitative analysis. We therefore assume that the
matter vector--like states decouple at an intermediate 
energy scale and study the conditions for obtaining
$\Lambda_h\approx1{\rm GeV}$.

With this assumptions the one--loop Renormalization Group
Equation (RGE) for the hidden group gauge coupling is given by
\beq
{1\over{\alpha_h(\mu)}}={1\over\alpha_0}-{1\over{2\pi}}
\left({1\over2}N_f-3N_C\right)\ln{M_h\over M_S}-{1\over{2\pi}}
\left(-3N_C\right)\ln{\Lambda_h\over M_h}
\label{intermediatemh}
\eeq
Writing $M_h=10^h{\rm GeV}$, and taking $\Lambda_h=1{\rm GeV}$;
$\alpha_h(\Lambda_h)=1$; $\alpha_0=1/24$, we obtain a relation
between $N_f$, $N_c$ and $h$,
\beq
{1\over2}N_f(16-h)=48N_C-{{46\pi}\over{\ln10}}
\label{nfnch}
\eeq
First we see that, since $0<h<16$ for any $N_C$ there will be solutions
to this equation, which depend on the number of flavors. This is
of course true, as for any $N_C$ we can add a number of flavor
multiplets that slow the evolution of the gauge couplings.
For larger $N_C$ we, of course, have to add more flavor 
multiplets. The more constraining framework then
has to be sought in the context of the realistic
superstring models. 

For $SU(2)$ from eq. (\ref{nfnch}) we see that if the intermediate 
scale $M_h$ is just an order of magnitude below the MSSM unification
scale, we need approximately 66 flavors in order for $\Lambda_h$
to be of order $1{\rm GeV}$, 16 flavors if $M_h\approx 10^{12}{\rm GeV}$
and 8 flavors if $M_h\approx 10^{8}{\rm GeV}$.  For $SU(3)$ we need
162 flavors for $M_h\approx10^{15}{\rm GeV}$, 40 flavors for
$M_h\approx10^{12}{\rm GeV}$, and 20 flavors for 
$M_h\approx10^{8}{\rm GeV}$. The number of needed flavors, of course, 
grows rapidly with increasing $N_c$. In the next section
we examine the feasibility of obtaining the needed spectrum
in realistic string models. However, we can already infer that
the desirable gauge group should have the smallest number
of colors, {\it i.e.} $SU(2)$. 

\section{String origins}\label{string}

In the previous section we showed that in order to get a hidden
sector which becomes strongly interacting at the GeV scale 
requires that the hidden gauge group has a small gauge content and
the existence of the hidden vector--like matter states at 
an intermediate energy scale. In this section we examine
whether these needed characteristics can be obtained from
realistic heterotic string models. From a purely field 
theoretic point of view we can of course construct a 
model with any number of colors and flavors, and a 
potential that will generate the required scales.
The resulting model may be rather contrived, but not
impossible to conceive. The more constraining framework
can therefore be sought in the context of string theory.
The class of string models that are of most interest in this
respect are those that can potentially reproduce the
observed physics of the standard particle model. It is
then of great interest to examine if such models can produce
a hidden sector with the characteristics that we discussed
in the previous sections. 

Examples of semi--realistic string models were constructed in
the orbifold and free fermionic formulations. The most realistic
models constructed to date are the free fermionic models which
utilize the NAHE\footnote{NAHE=pretty in Hebrew. The NAHE
set was first employed by Nanopoulos, Antoniadis, Hagelin and Ellis
in the construction of the flipped $SU(5)$ heterotic--string model
\cite{flipped}. Its vital role in the realistic free fermionic models has been
emphasized in ref.~\cite{nahe}.} set of boundary condition basis vectors
\cite{nahe,slm}. These constructions naturally give rise to three generation
models with the standard $SO(10)$ embedding of the Standard 
Model spectrum\footnote{It is interesting to note that among the perturbative
heterotic--string orbifold models the free fermionic models
are the only ones which have yielded three generations with the 
canonical $SO(10)$ embedding}. Furthermore, one of the generic features
of semi--realistic string vacua is the existence of 
numerous massless states beyond the MSSM spectrum, some
of which carry fractional electric charge and hence must decouple
from the low energy spectrum. Recently, and for the first
time since the advent of string phenomenology, it was demonstrated
\cite{cfn} in the FNY free fermionic model \cite{fny,fcp},
that free fermionic models 
can also produce models with solely the MSSM states in the light
spectrum.

The realistic free fermionic models are defined in terms of a set of
boundary condition basis vectors for all the world--sheet fermions,
and the one--loop GSO amplitudes \cite{fff}. The physical massless states
in the Hilbert space are obtained by acting on the vacuum
with bosonic and fermionic operators and by applying the 
generalized GSO projections. The basis is constructed in
two stages. The first stage consists of the NAHE set \cite{nahe,slm}, 
which is a set of five boundary condition basis 
vectors, $\{{{\bf 1},S,b_1,b_2,b_3}\}$. The gauge group after the
NAHE set is $SO(10)\times SO(6)^3\times E_8$ with $N=1$ space--time
supersymmetry. The space--time vector bosons that generate the gauge group 
arise from the Neveu--Schwarz (NS) sector and from the sector
$\zeta\equiv {\bf 1}+b_1+b_2+b_3$. The NS sector produces the generators of 
$SO(10)\times SO(6)^3\times SO(16)$. The sector 
$\zeta$ produces the spinorial $\bf 128$ of $SO(16)$ and completes the hidden 
gauge group to $E_8$. 
The sectors $b_1$, $b_2$ and $b_3$ 
produce 48 spinorial $\bf 16$'s of $SO(10)$, sixteen from each sector $b_1$, 
$b_2$ and $b_3$. 

The second stage of the basis construction consist
of adding three additional basis vectors to the NAHE set
typically denoted by $\{\alpha,\beta,\gamma\}$. 
Three additional vectors are needed to reduce the number of generations 
to three, one from each sector $b_1$, $b_2$ and $b_3$. 
At the same time the additional boundary condition basis vectors
break the gauge symmetries of the NAHE set.
The $SO(10)$ symmetry is broken to one of its subgroups.
The flavor $SO(6)^3$ symmetries are broken to product of $U(1)$'s,
and the hidden $E_8$ is broken to one of its subgroups.
In addition to the spin one and two multiplets, the Neveu--Schwarz 
(NS) sector produces three pairs of electroweak doublets,
$\{h_1, h_2, h_3, {\bar h}_1, {\bar h}_2, {\bar h}_3\}$,
three pairs of $SO(10)$ singlets with $U(1)$ charges,
$\{\Phi_{12},\Phi_{23},\Phi_{13},{\bar\Phi}_{12},
{\bar\Phi}_{23}, {\bar\Phi}_{13}\}$,
and three singlets of the entire four dimensional gauge group,
$\{\xi_{1},\xi_2, \xi_3\}$.
This generic structure is common to a large number of three generation
models which differ in their detailed phenomenological
characteristics. The analysis of the models proceeds by analyzing the
cubic level and higher order terms in the superpotential and
by imposing that the string vacuum preserves $N=1$ space--time
supersymmetry. By studying specific models in this fashion,
it was demonstrated that these models can potentially
reproduce the fermion mass spectrum, and also produce
models with solely the MSSM spectrum in the observably charged
sector. 

The fact that the free fermionic models produce models
that look tantalizingly realistic renders the search
for possible signatures beyond the observed spectrum
much more appealing. In this paper we focus on the 
possibility of the strongly interacting dark matter.

The basis vectors $\{\alpha, \beta, \gamma\}$, which break the 
observable gauge group, also break the hidden $E_8$ gauge
group to one of its subgroups. This is a necessary
consequence of the perturbative string consistency conditions, 
{\it i.e.} of modular invariance. The resulting hidden gauge
groups which arise depend on the specific models and are
quite varied. We will comment more on this below.  
In addition to the generic spectrum from the NS sector
and the sectors $b_1$, $b_2$ and $b_3$, outlined above,
the models typically also contain additional massless 
states, which arise from the basis vectors which extend
the NAHE set. The three sectors $b_j+2\gamma$ produce hidden matter 
states that fall into the $\bf16$ representation of the
$SO(16)$ subgroup of the hidden $E_8$, decomposed
under the final $E_8$. These states are $SO(10)$ singlets
but are charged under the horizontal $U(1)$ symmetries. 
In addition, vectors that are combinations of the NAHE set basis vectors
and of the basis vectors $\{\alpha,\beta,\gamma\}$, produce additional
massless sectors which break the $SO(10)$ symmetry explicitly.
Some of these states are Standard Model singlets and 
can therefore also remain in the light spectrum. 

The discussion above summarizes the general structure of the 
realistic free fermionic models. Our next task is then 
to make a survey of several models and to examine whether
there exist, if any, models that can produce the
characteristics desired for the self--interacting dark matter. 
As we discussed in section \ref{rge} to obtain $\Lambda_h\sim1{\rm GeV}$
we need a non--Abelian gauge group with small gauge content
and matter spectrum at an intermediate mass scale.
The prerequisite from the string model perspective is that
the hidden $E_8$ gauge group is broken to a sufficiently small 
factor. Most favorably the hidden sector should contain an 
$SU(2)$ or $SU(3)$ gauge groups. The second condition is that
there should be a sufficient number of hidden matter multiplets
to slow down the evolution of the gauge coupling. 

The revamped flipped $SU(5)$ model of ref. \cite{flipped}
produces a hidden sector with $SO(10)\times SO(6)$ gauge
group. The model contains 5 multiplets in the 10 vectorial
representation of the hidden $SO(10)$; 5 multiplets in the 
6 vectorial representation of the hidden $SU(4)$ and
5 multiplets in the $(1,4)\oplus (1,{\bar 4})$
representations. For $SO(10)$ we find that even if we
assume that all the matter states remain massless, 
the theory becomes strongly interacting at $\Lambda_h\sim5\cdot
10^{12}{\rm GeV}$ (taking $M_S=10^{16}{\rm GeV}$). For
$SU(4)$ we have that taking all the spectrum
to remain massless gives $b_{SU(4)_h}=-2$, which may  
result in a sufficiently low scale for the hidden $SU(4)$. 
However, in the revamped flipped $SU(5)$ model the $4+\bar4$
states carry fractional electric charge and therefore 
must either decouple or the theory must confine at 
a much higher scale \cite{flipped,crypt}. 

We next turn to the $SO(6)\times SO(4)$ model of
ref. \cite{alr}. In this model the hidden gauge group
is $SU(8)$ and there are 5 multiplets in the $8+\bar8$ 
representations. This again yields $\Lambda_h\sim5\cdot10^{12}
{\rm GeV}$ (taking $M_S\sim 10^{16}{\rm GeV}$, even if we
assume that all of the matter states remain massless. 

Next we turn to the case of the string Standard--like Models \cite{fny,slm,eu}.
These models represent the most interesting possibilities
for the strongly interacting dark matter for the following
reason. As we discussed above it is desirable to have
a hidden gauge sector that contains small group factors, 
like $SU(2)$ and $SU(3)$. In the string standard--like models
the observable gauge group is broken by two
subsequent basis vectors. The modular invariance constraints
then impose that similarly the hidden gauge group in these
models has to be broken by the same two basis vectors.
This means that in the string standard--like models
small hidden group factors can indeed naturally arise. 
The same argument also applies to the left--right symmetric
models of ref. \cite{cfs}. On the other hand in the $SU(5)\times
U(1)$ or $SO(6)\times SO(4)$ type models, the observable $SO(10)$
gauge group is broken by a single basis vector. Consequently,
the hidden gauge group in these models contains larger 
group factors. 

Turning then to the string standard--like models
we find that indeed $SU(2)$ and $SU(3)$ hidden group factors
frequently arise. For example, in the model of
ref. \cite{eu} the hidden gauge group is $SU(5)\times SU(3)\times U(1)^2$.
There exist 8 multiplets in the $3+{\bar3}$ representations,
producing $b_{SU(3)_h}=-1$. Of those five carry fractional electric
charge and therefore must decouple at a high scale. 
If we take the scale $M_I$ at which all the hidden matter fields
decouple as in section \ref{rge}, then requiring $\Lambda_h\sim1{\rm
GeV}$ imposes $M_h\sim 10^6{\rm GeV}$ for the intermediate
energy scale. If on the other hand we assume that the
electrically neutral states remain light down to $\Lambda_h$
and that the fractionally charged states decouple at the 
scale $M_I$ then we find that $M_I\sim 10^{11}{\rm GeV}$,
which seems more reasonable. All in all this qualitative analysis
suggests that models with a hidden $SU(3)$ group 
factor may have enough flexibility to allow a small
$\Lambda_h$. 

Next we turn to the model of ref. \cite{fny}. The hidden
gauge group in this model is $SU(3)\times SU(2)\times SU(2)\times 
U(1)^4$. Each hidden $SU(2)$ gauge group contains 10
multiplets in the fundamental representation, which are 
all electrically neutral. If we assume that all the matter states
decouple at an intermediate scale $M_I$, then imposing 
$\lambda_h\sim 1{\rm GeV}$ we obtain $M_h\sim10^{11}{\rm GeV}$. 
This model therefore demonstrates that hidden gauge
group with a low confining scale and with matter which decouples
at a much higher scale may indeed arise in realistic
string models. 

\section{Discussion}

The proposal of the self--interacting dark matter is a new interesting 
development which can help to reconcile computer--simulated features of 
galactic substructures with observations. Whether the self-interacting
dark matter represents a considerable improvement over a more conventional 
cold dark matter scenario is, of course, an open question, which we are 
not trying to address here. Nevertheless, it is interesting to explore a 
possibility of having self--interacting dark matter within a particle physics 
context. 

In this paper we argued that a naturally self--interacting dark matter
can arise from the hidden sector gauge group. If this gauge group is not 
Higgsed, it will enter a strongly--interacting regime at some scale 
$\Lambda_h$, hadronize and develop a mass gap. 
A lightest particle, with the mass presumably of the order $\Lambda_h$,
will be stable and will have an elastic cross section roughly on the order of 
$\Lambda_h^{-2}$. If this scale $\Lambda_h$ is 1 GeV or smaller, these 
glueballs may fit into self--interacting dark matter criterion, put forward 
by Spergel and Steinhardt. The connection between the hidden and visible 
sectors may occur due to exotic heavy matter states, charged under the SM and 
hidden gauge groups, which decouple at a very high scale $M_S$, presumably 
comparable to unification scale. As a result, 
 this connection is mediated 
by dimension 8 operators (dimension 6, if the SM Higgs has couplings to 
the exotic matter states). 
This induces a decay of exotic glueballs, suppressed by the eighth (fourth) 
power of the heavy scale. Consequently, 
if the condition on the self--interacting dark matter is satisfied, 
this suppression makes the life time of exotic glueballs to be much 
larger than the present age of the Universe.

The extreme smallness of the coupling between the visible and hidden sector at 
low energies poses certain difficulties in explaining the cosmological 
abundance of exotic glueballs, close to a required value. 
Indeed, since there is no ways of diluting the energy stored in the hidden 
sector through the decay into the SM particles, we 
conclude that the visible and hidden sectors have never been in thermal 
equilibrium. Consequently, we have to assume that the energy was stored
in the hidden sector soon after inflation, either due to a direct coupling 
of the inflaton into exotic gluons or through the annihilation of the 
visible sector particles into exotic gluons. The efficiency of a latter 
process is governed again by the same effective operators, which connect 
the visible and hidden sectors.
This process puts an upper limit on the reheating 
temperature, $T_{RH}\leq 10^{-4} M_S$ for dimension 6 operators and 
$T_{RH}\leq 10^{-2} M_S$ for dimension 8. 

Choosing a specific gauge group, we can connect the two scales, $M_S$ and 
$\Lambda_h$. If we insist on the unification of couplings at $M_S$, 
the condensation scale $\Lambda_h$ is many orders of magnitude larger than 
desirable value of 1 GeV even for the minimal gauge group $SU(2)$. 
This problem can be cured only if one assumes a number of matter-like 
thresholds at some intermediate scale which would reduce the initial 
value of the coupling constant. 

The cold dark matter candidates, like the axion and the neutralino,
are well motivated by theoretical considerations. Does the
stable hidden glueball have a similar appealing theoretical 
motivation? We believe that the the answer is indeed affirmative.
The existence of the hidden sector is a natural consequence of
string theory, the only prevalent theory that at present
offers a viable framework for quantum gravity. 
Realistic string models that reproduce the general 
structure of the Standard Model spectrum and 
have the potential of explaining its detailed features,
also produce a hidden sector with the general characteristics 
that we assumed in this paper. Namely, a non--Abelian
hidden gauge group with matter states in vector--like representations.
It is then most intriguing that the models that come the closest
to being fully realistic also produce the hidden gauge group
with small gauge content, as is required if the hidden
gauge group is to confine at the hadronic scale. Further
exploration to reveal whether other classes of string models 
\cite{othermodels} can produce a hidden sector with similar
characteristics are of enormous interest. Finally, at the turn of the new
millenium it seems that a burning question in physics
is: What is the universe made of? The answer to this question,
whether in the direction advocated in this paper or otherwise,
will have profound implications on our basic understanding
of fundamental physics. 

\bigskip
\medskip
\leftline{\large\bf Acknowledgments}
\medskip
We are pleased to thank Keith Olive and Misha Voloshin for useful discussions.
This work was supported in part by the U.S. DOE 
Grant No.\ DE-FG-02-94-ER-40823 and a PPARC advanced fellowship (AEF).



\bibliographystyle{unsrt}

\end{document}